Magnetic and transport properties of the new antiferromagnetic Kondo-lattice CeNiBi$_2$

M. H. Jung* and A. H. Lacerda

*National High Magnetic Field Laboratory, Los Alamos National Laboratory, MS E536 Los Alamos, NM 87545*

T. Takabatake

*Department of Quantum Matter, ADSM, Hiroshima University, Higashi-Hiroshima 739-8526, Japan*

We report results of the first studies on the magnetic and transport properties of a new material CeNiBi$_2$. The magnetic susceptibility exhibits a sharp peak at $T_N$ = 6K, indicating an antiferromagnetic phase transition. This antiferromagnetic order below $T_N$ is confirmed by magnetization measurement, which displays a metamagnetic-like transition at $H_m$ = 5 T. Both low-temperature susceptibility and high-field magnetization are suggestive of strong crystalline-electric-field effect in CeNiBi$_2$. The electrical resistivity shows the presence of Kondo and crystal-field effects with a sharp drop below $T_N$ due to the antiferromagnetic ordering. This sharp drop below $T_N$ in the electrical resistivity is suppressed slightly to higher temperatures by an applied magnetic field to 18 T. With increasing magnetic field, the slope of magnetoresistance changes from positive to negative, being indicative of the transition to a ferromagnetic state.

*To whom correspondence should be addressed. E-mail: mhjung@lanl.gov

Ternary intermetallic compounds containing the rare-earth metal, transition metal, and nonmetallic or semimetallic element are the subject of continuous interest because they show a wide variety of exotic physical properties: Kondo effect, intermediate valence, and coexistence of heavy fermion behavior with various magnetic orderings (see ref. 1 for a review). Searching in the ternary systems of Ce-Ni-Bi, a new phase of $CeNiBi_2$ has recently been found and studied carefully. X-ray powder diffraction patterns indicated that $CeNiBi_2$ crystallizes in the tetragonal $ZrCuSi_2$-type structure (space group P4/nmm) with the lattice parameters of $a=4.54(2)$Å and $c=9.63(8)$ Å, in good agreement with that reported previously.[2] The present paper reports the first results of magnetic susceptibility, magnetization, electrical resistivity, and magnetoresistance measurements. It was found that $CeNiBi_2$ is an antiferromagnetic material with the Néel temperature of $T_N = 6$ K and shows the presence of Kondo and crystal-field effects.

The temperature dependence of magnetic susceptibility $\chi(T)$ and its inverse $\chi^{-1}$ measured in a field of 0.1 T is shown in Fig. 1 with the inset clarifying the low-temperature behavior. A peak at 6 K in the susceptibility indicates an antiferromagnetic ordering of the Ce moments. The high-temperature susceptibility does not obey the simple Curie-Weiss law $\chi = N\mu_{eff}^2/3k_B(T-\theta_P)$, but it could be fitted to a modified Curie-Weiss law which is suitable for materials having a large temperature-independent susceptibility: $\chi = N\mu_{eff}^2/3k_B(T-\theta_P) + \chi_o$, where $\chi_o$ represents the temperature-independent part of the magnetic susceptibility, including the core-electron diamagnetism, the Pauli paramagnetism, and Van Vleck terms. This approach works well for materials like Kondo or valence fluctuating systems, such as $Ce_2Ni_2Cd$,[3] $CeIrGe$,[4] $CeRuGe_3$,[5] and $Ce_2CoSn_2$.[6] The data are very well fitted from 300 K down to 50 K in

CeNiBi$_2$ (solid line in Fig. 1). We obtain the constant value of $\chi_o$ = 0.00791 emu/mol, the effective magnetic moment of $\mu_{eff}$ = 2.83 $\mu_B$/Ce, and the paramagnetic Curie temperature of $\theta_P$ = – 27.7 K. The observed value of $\mu_{eff}$ is much higher than that one would expect for a free Ce$^{3+}$ ion (2.54 $\mu_B$/Ce). This might indicate that the magnetic moments of Ce ions are well localized in the compound. The negative sign of $\theta_P$ is suggestive of a tendency toward antiferromagnetic correlation between the Ce moments at high temperatures. The deviation from the modified Curie-Weiss behavior below 50 K could be attributed to the crystalline-electric-field (CEF) effect on the ground state.

The isothermal magnetization $M(H)$ at 1.5 K is shown in Fig. 2. Neither hysteresis nor remanence was observed on increasing and decreasing applied magnetic field. The magnetization increases rapidly at low fields to 0.5 T, increases slowly to 2 T, exhibits a maximum in the derivative of the magnetization at 5 T, and shows partial saturation at higher fields. No complete saturation was observed up to 15 T. The value of the magnetic moment at 15 T reaches a value of about 0.8 $\mu_B$/Ce, which is much less than the value expected for the magnetic susceptibility of CeNiBi$_2$. Both the smaller value of the saturation magnetization and the low-temperature susceptibility behavior could be indicative of important CEF effect in this compound. Inelastic neutron scattering studies are further required in order to understand the magnetization in the presence of CEF states. The plot of the derivative of the magnetization d$M$/d$H$ exhibits a broad peak at 5 T, which might be attributed to a metamagnetic-like transition, probably due to a spin-flip type.

Figure 3 displays the temperature dependence of electrical resistivity $\rho(T)$ measured in a transverse configuration ($H \perp I$) at various magnetic fields to 18 T. The zero-field resistivity shows a broad shoulder at 100 K and a sharp peak at $T_N$ = 6 K. The former is likely to be associated with the interplay between Kondo and CEF effects as found for other Ce-based compounds such as $CeNiX_2$ (X = Si, Ge, Sn)[7] and $CeTGe_2$ (T = Ni, Rh, Ir).[8,9] The latter is attributed to the antiferromagnetic ordering observed at $T_N$ = 6 K in the magnetic susceptibility. With increasing magnetic field, however, the peak at $T_N$ is weakened and shifts to lower temperature, then a small kink appears in fields $H \geq 10$ T. For $H$ = 18 T, the anomaly was observed at 4.5 K (the inset of Fig. 3).

The normalized transverse magnetoresistance, $\Delta\rho/\rho_o = [\rho(H) - \rho(0)]/\rho(0)$, is plotted as a function of the applied magnetic field at different temperatures in Fig. 4. At 5 K below $T_N$, the magnetoresistance is initially positive and then turns negative at higher fields, making a maximum at $H_m$ = 4.2 T. This maximum moves to higher fields as the temperature is increased to $T$ = 10 K, where it is at $H_m$ = 6 T. The positive magnetoresistance at low fields can be understood by taking account of an enhancement of spin-disorder scattering as the antiferromagnetic state is changed into a field-induced ferromagnetic state. The change of the slope of $\Delta\rho/\rho_o$ can be interpreted in light of the spin-polarized effect. Since the magnetic scattering in a ferromagnetic ground state is much weaker than that in an antiferromagnetic ground state, the high-field limit of the normalized magnetoresistance should be negative. In other words, the negative magnetoresistance at high magnetic fields is a result of the strong reduction of scattering by the ferromagnetic alignment of Ce magnetic moments. This result is similar to that observed in $CeCoGe_3$,[10] but it is different from that found for $CeRh_2Ge_2$ where the field of $H_m$ shifts to lower fields as the temperature is increased.[11] The magnetoresistance at 30 K is positive over all the magnetic fields.

Among the rare-earth ternary intermetallic compounds, CeNiBi$_2$ is a new antiferromagnetically ordered Kondo-like compound with the Néel temperature $T_N$ = 6 K. The magnetic susceptibility does not obey the simple Curie-Weiss law but obeys a modified Curie-Weiss law. Both low-temperature susceptibility and high-field magnetization are reflective of important CEF effect. The presence of a broad shoulder at 100 K in the electrical resistivity is an indication of the interplay between Kondo and CEF effects. These features are similar to other intermetallics CeNiX$_2$,[7] in which CeNiSi$_2$ undergoes valence fluctuation at high temperatures and spin fluctuation at low temperatures and both CeNiGe$_2$ and CeNiSn$_2$ exhibit antiferromagnetic phase transitions accompanied by a transition to field-induced ferromagnetic state. The metamagnetic-like transition from the antiferromagnetic state to a field-induced ferromagnetic state in CeNiBi$_2$ is supported by the observation of the maximum at 5 T in the derivative of the magnetization and the change of the slope of the magnetoresistance at temperatures below $T_N$. Besides, we conclude that the magnetotransport of CeNiBi$_2$ depends strongly on an applied magnetic field, i. e., magnetic alignment of Ce moments and/or magnetic scattering mechanism.

Figure captions

Fig. 1. Temperature dependence of magnetic susceptibility $\chi(T)$ measured in a field of 0.1 T for CeNiBi$_2$. The solid line represents the best fit of the modified Curie-Weiss law (see text). The inset shows the low-temperature behavior.

Fig. 2. Magnetization $M(H)$ measured at $T = 1.5$ K for CeNiBi$_2$. The inset shows the differential magnetization d$M$/d$H$ as a function of applied magnetic field.

Fig. 3. Temperature dependence of electrical resistivity $\rho(T)$ for CeNiBi$_2$ in zero and 18 T applied field in the transverse configuration ($H \perp I$). The inset shows the low-temperature data at various magnetic fields of 0, 5, 10, and 18 T.

Fig. 4. Normalized magnetoresistance $\Delta\rho/\rho_o = [\rho(H) - \rho(0)]/\rho(0)$ measured at various temperatures 5, 10, 30 K for CeNiBi$_2$ in the transverse configuration ($H \perp I$).

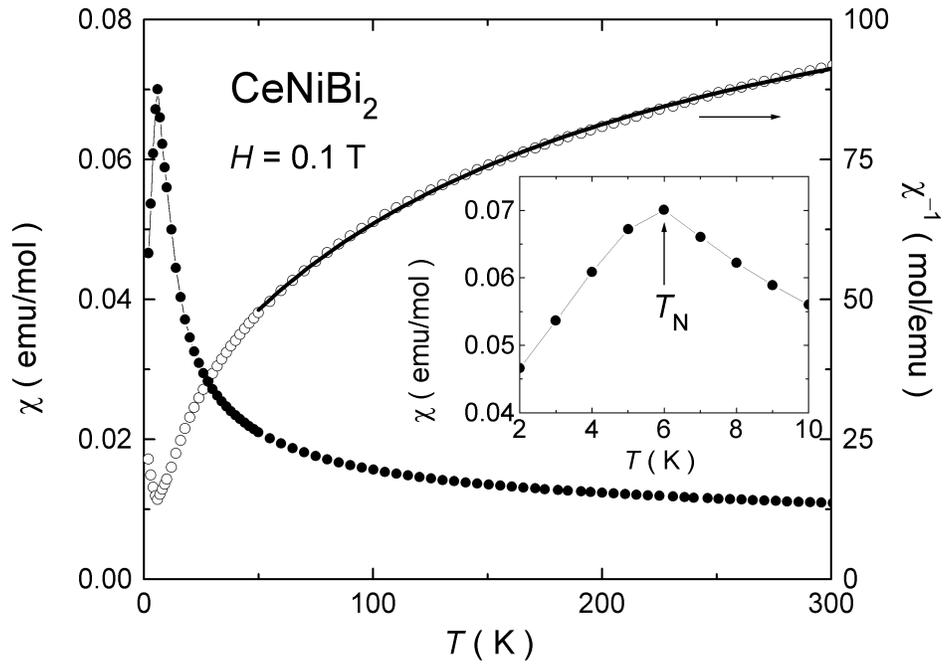

Fig. 1. M. H. Jung et al.

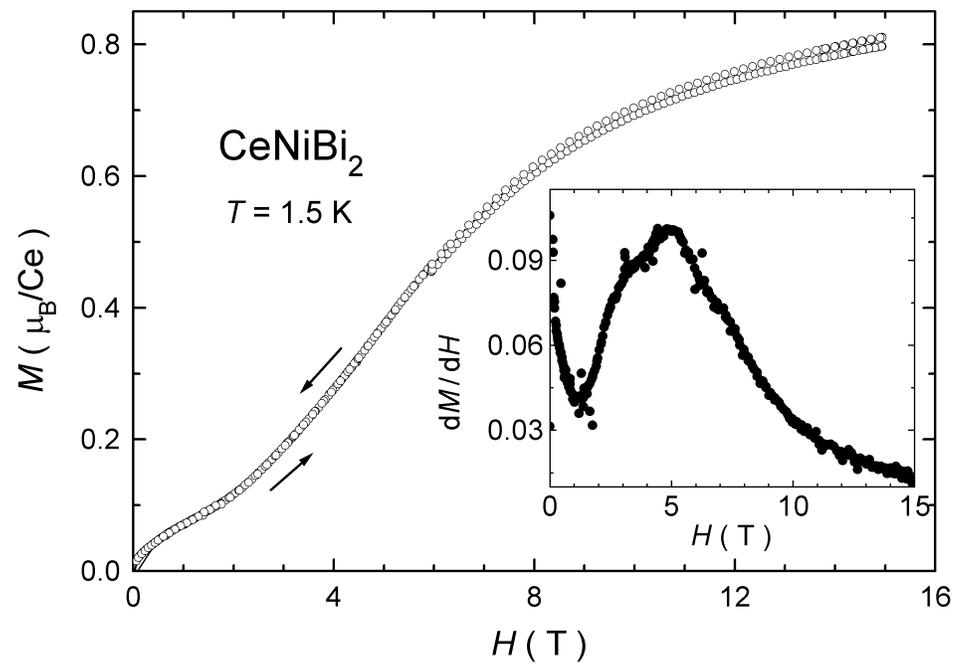

Fig. 2. M. H. Jung et al.

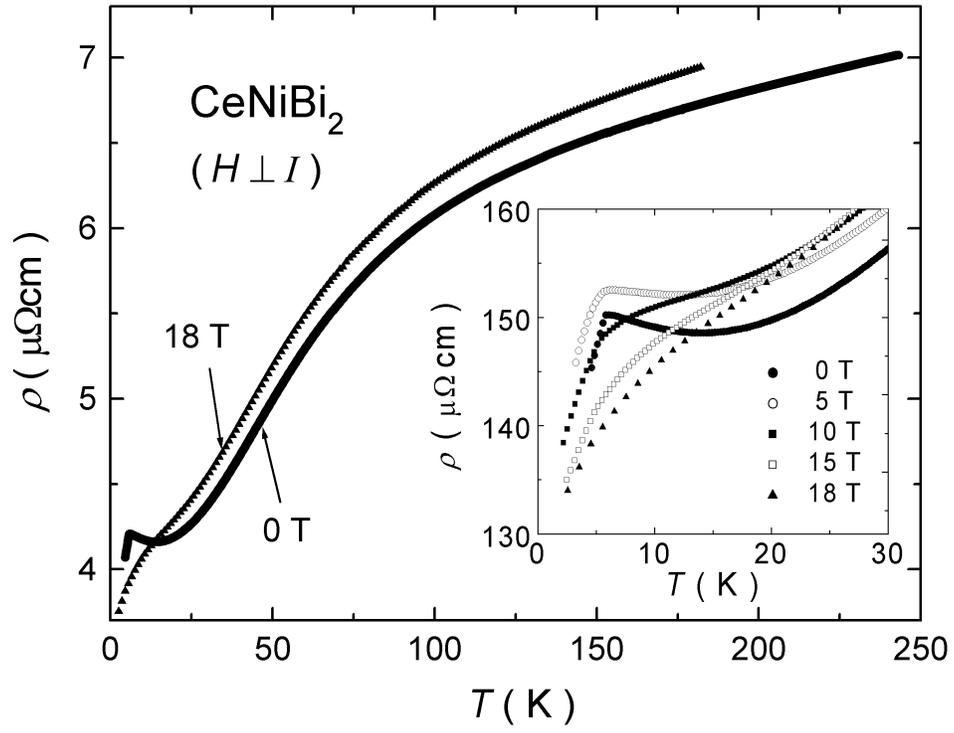

Fig. 3. M. H. Jung et al.

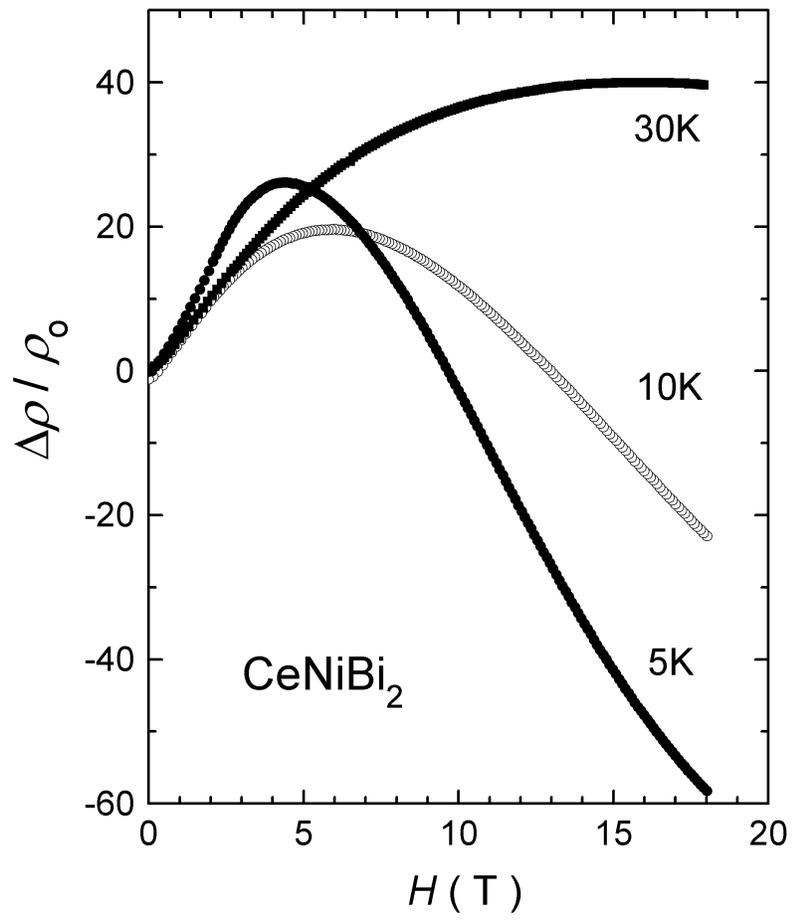

Fig. 4. M. H. Jung et al.